\documentclass[amsfonts, prl,floats,aps,superscriptaddress,twocolumn,narrowtext]{revtex4}
\usepackage{amsmath}
\usepackage{bm}
\usepackage{exscale}
\usepackage[mathscr]{eucal}
\usepackage{epsfig}

\begin{document}

\title{Screening properties of a two-component charge Bose gas}

\author{K. K. Lee}
\affiliation{IRC in Superconductivity, Cavendish Laboratory,
University of Cambridge, Cambridge, CB3 0HE, United Kingdom}

\begin{abstract}
The excitation spectrum of a two-component Bose-Einstein
condensate has been discussed by Ref \cite{alex2comp} using the
Bogoliubov equations. Here we make the additional contributions of
the linear response function and dielectric function for a
two-component CBG in the condensation phase as well as the
confirmation of the plasmon excitation spectrum using the
Gross-Pitaevskii (G-P) equations.
\end{abstract}

\maketitle
\section{Introduction} \label{sect:thefirst}

The properties of a two-component Bose-Einstein condensate were
discussed by Alexandrov and Kabanov \cite{alex2comp}. There the
Bogoliubov theory was extended for a two-component, non-converting
three-dimensional charge or neutral Bose gas. Here we make the
additional contributions of the linear response function and
dielectric function for a two-component CBG in the condensation
phase ($\mu_j=0$) as well as the confirmation of the plasmon
excitation spectrum using the Gross-Pitaevskii (G-P) equations.

Studies of the Bose condensation temperature in the cuprates
\cite{alexfree, alexbook} and also recent experiments on BEC in
$^{87}Rb$ atoms in two different hyperfine states \cite{hagley}
gives the main motivation. Such a model could apply for the
cuprates since the two components may describe bipolarons
travelling in either the x- or y-direction.

\section{Optical and acoustic plasmon modes} \label{sect:thesecond}
We give an outline of the main results from studies by Alexandrov
et al \cite{alex2comp}. In an external vector potential
$\bf{A}(\bf{r},t)$ and a scalar potential $U_j(\bf{r},t)$, the
Hamiltonian for a two-component (j=1,2) Bose gas is given as
\begin{widetext}
\begin{equation}
H=\sum_{j=1,2}\int d \bm{r} \psi^{\dagger}_j(
\bm{r})\left[-\frac{( \bm{\nabla}-iq_j\bm{A}(\bm{r},t))^2}{2m_j}
+U_j(\bm{r},t)-\mu_j\right]\psi_j(\bm{r})+\frac{1}{2}\sum_{j,j'}\int
d\bm{r}\int d\bm{r}' V_{jj'}(\bm{r}-\bm{r}')
\psi^{\dagger}_j(\bm{r})
\psi_j(\bm{r})\psi^{\dagger}_{j'}(\bm{r}') \psi_{j'}(\bm{r}'),
\end{equation}
\end{widetext}
where $q_j$ and $m_j$ are the charge and mass of the $j^{th}$
Bosons. Provided the two body interaction $V_{jj'}(\bm{r})$ is
weak, one may assume the ground state is macroscopically occupied.
Hence one can apply a Bogoliubov displacement transformation
\begin{equation}
\psi(\bm{r},t)=\phi(\bm{r},t)+\tilde{\psi}(\bm{r},t)
\end{equation}
and consider the excitations as fluctuations from the ground
state. Furthermore one then applies a Bogoliubov linear
transformation for the fluctuations
\begin{equation}
\tilde{\psi}(\bm{r},t)=\sum_n u_{nj}(\bm{r},t)(\alpha_n+\beta_n)
+v^*_{nj}(\bm{r},t)(\alpha^{\dagger}_n+\beta^{\dagger}_n).
\end{equation}
One obtains the Bogoliubov-de Gennes (BdG) equations:
\begin{widetext}
\begin{equation}
i\frac{d}{dt}u_j(\bm{r},t)-\hat{h}u_j(\bm{r},t)=\sum_{j'}\int
d\bm{r}'V_{jj'}(\bm{r}-\bm{r}') [
|\phi_{j'}(\bm{r}',t)|^2u_{j}(\bm{r},t)
+\phi^*_{j}(\bm{r}',t)\phi_{j'}(\bm{r},t)u_{j'}(\bm{r}',t)
+\phi_{j}(\bm{r}',t)\phi_{j'}(\bm{r},t)v_{j'}(\bm{r}',t) ]
\end{equation}
\end{widetext}
and
\begin{widetext}
\begin{equation}
-i\frac{d}{dt}v_j(\bm{r},t)-\hat{h}^*v_j(\bm{r},t)=\sum_{j'}\int
d\bm{r}'V_{jj'}(\bm{r}-\bm{r}') [
|\phi_{j'}(\bm{r}',t)|^2v_{j}(\bm{r},t)
+\phi_{j}(\bm{r}',t)\phi^*_{j'}(\bm{r},t)v_{j'}(\bm{r}',t)
+\phi^*_{j}(\bm{r}',t)\phi^*_{j'}(\bm{r},t)u_{j'}(\bm{r}',t) ],
\end{equation}
\end{widetext}
where
\begin{equation}
\sum_n [u_{nj}(\bm{r},t)u^*_{nj}(\bm{r}',t)
+v_{nj}(\bm{r},t)v^*_{nj}(\bm{r}',t)]=\delta(\bm{r}-\bm{r}')
\end{equation}
and
\begin{equation}
\hat{h}_j=-\frac{(\bm{\nabla}-iq_j\bm{A}(\bm{r},t))^2}{2m_j}
+U_j(\bm{r},t)-\mu_j.
\end{equation}
Using these BdG equations one finds the elementary excitations for
a two-component Bose condensate with arbitrary interactions:
\begin{widetext}
\begin{equation}
E_{1,2}(\bm{k})=\frac{1}{2^{1/2}}\left(
\varepsilon^2_1(\bm{k})+\varepsilon^2_2(\bm{k})\pm\sqrt{
[\varepsilon^2_1(\bm{k})-\varepsilon^2_2(\bm{k})]^2+
\frac{4k^4}{m_1m_2}W^2_{\bm{k}}n_1n_2}\right)^{1/2}
\end{equation}
\end{widetext}
where we assume no external fields and that the
excitation coherence factors are plane waves
\begin{eqnarray}
u_{\bm{k},j}(\bm{r},t)=u_{\bm{k},j}e^{i(\bm{k}.\bm{r}-E_{\bm{k}}t)}
\\
v_{\bm{k},j}(\bm{r},t)=v_{\bm{k},j}e^{i(\bm{k}.\bm{r}-E_{\bm{k}}t)}.
\end{eqnarray}
Also we have used the definitions
\begin{eqnarray}
\varepsilon_1(\bm{k})=\sqrt{\frac{k^4}{4m^2_1}+\frac{k^2V_{\bm{k}}n_1}{m_1}}
\\
\varepsilon_2(\bm{k})=\sqrt{\frac{k^4}{4m^2_2}+\frac{k^2U_{\bm{k}}n_2}{m_2}}
\end{eqnarray}
and where $V_{\bm{k}},U_{\bm{k}}$ and $W_{\bm{k}}$ are the Fourier
components of the interactions $V_{11}(\bm{r}), V_{22}(\bm{r})$
and $V_{12}(\bm{r})=V_{21}(\bm{r})$ respectively. This result
implies there are two branches, acoustic and optical. The acoustic
branch is quadratic in k and means that the condition for
superfluidity is not met (Landau criterium). However such a
problem can be avoided by considering more than one interaction
potential, for example a hard-core and a long-range potential.
Therefore a two-component bipolaron gas may still be a superfluid.

\section{Linear response and the dielectric function} \label{sect:theforth}
Here we derive the linear response function for a two-component
CBG. From this we obtain the dielectric function and examine its
behaviour. In the process we find the same excitation spectrum
found earlier using the Bogoliubov equations of the previous
section.

The G-P equation in the condensed phase is written:
\begin{widetext}
\begin{equation}
i\frac{d}{dt}\psi_{0j}(\bm{r},
t)=-\frac{(\bm{\nabla}-ie^*\bm{A})^2}{2m_j}\psi_{0j}(\bm{r},
t)+\sum_{j'}\int
d\bm{r}'V_{jj'}(\bm{r}-\bm{r}')|\psi_{0j'}(\bm{r}',
t)|\psi_{0j}(\bm{r}, t)
\end{equation}
\end{widetext}
We allow for small fluctuations of the order parameter as
\begin{equation}
\psi_{0j}(\bm{r}, t)=\sqrt{n_{0j}}+\phi_j(\bm{r}, t)
\end{equation}
and keeping terms linear in $\bm{A}$ and fluctuations
$\phi_j(\bm{r}, t)$ and assuming we have a homogeneous background
charge:
\begin{equation}
\int d\bm{r}'V(\bm{r}')=0,
\end{equation}
we obtain
\begin{widetext}
\begin{equation}
i\frac{d}{dt}\phi_{j}(\bm{r},
t)=-\frac{\bm{\nabla}^2}{2m_j}\phi_j(\bm{r}, t) +
\frac{e^*\sqrt{n_{oj}}}{2m_j}\bm{\nabla}\cdot\bm{A}(\bm{r}, t)
+\sum_{j'}\int
d\bm{r}'V_{jj'}(\bm{r}-\bm{r}')\sqrt{n_{oj}n_{oj'}}\left[\phi_{j'}(\bm{r}',
t)+\phi^*_{j'}(\bm{r}', t)\right].
\end{equation}
\end{widetext}
By Fourier transformation this becomes
\begin{widetext}
\begin{equation}
\omega\phi_{j}(\bm{q},\omega)
=\frac{q^2}{2m_j}\phi_j(\bm{q},\omega) -
\frac{e^*\sqrt{n_{oj}}}{2m_j}\bm{q}\cdot\bm{a}(\bm{q},\omega)
+\sum_{j'}V_{jj'}(\bm{q},\omega)\sqrt{n_{oj}n_{oj'}}
\left[\phi_{j'}(\bm{q},\omega)+\phi^*_{j'}(-\bm{q},-\omega)\right].
\end{equation}
\end{widetext}
The complex conjugate is easily found to be
\begin{widetext}
\begin{equation}
-\omega\phi^*_{j}(-\bm{q},-\omega)
=\frac{q^2}{2m_j}\phi^*_{j}(-\bm{q},-\omega) +
\frac{e^*\sqrt{n_{oj}}}{2m_j}\bm{q}\cdot\bm{a}(-\bm{q},-\omega)
+\sum_{j'}V_{jj'}(\bm{q},\omega)\sqrt{n_{oj}n_{oj'}}
\left[\phi_{j'}(\bm{q},\omega)+\phi^*_{j'}(-\bm{q},-\omega)\right]
\end{equation}
\end{widetext}
where
\begin{equation}
\int
d\bm{r}'V(\bm{r}-\bm{r}')e^{-\bm{q}\cdot(\bm{r}-\bm{r}')}=V(\bm{q})
\end{equation}
and $\bm{a}(\bm{q},\omega)=\bm{a}^*(-\bm{q},-\omega)$ is the
fourier transform of the external vector potential. The 4
equations can be written in a matrix form as

\begin{widetext}
\begin{equation}
\left( \begin{array}{c} \frac{e^*\sqrt{n_{01}}}{2m_1} \\\\
\frac{e^*\sqrt{n_{02}}}{2m_2} \\\\
\frac{-e^*\sqrt{n_{01}}}{2m_1} \\\\
\frac{-e^*\sqrt{n_{02}}}{2m_2}
\end{array} \right)\bm{q}\cdot\bm{a}=
\left( \begin{array}{cccc} \frac{q^2}{2m_1}+V_{11}n_{01}-\omega &
V_{12}\sqrt{n_{01}n_{02}} & V_{11}n_{01} &
V_{12}\sqrt{n_{01}n_{02}}
\\\\
V_{21}\sqrt{n_{02}n_{01}} & \frac{q^2}{2m_2}+V_{22}n_{02}-\omega &
V_{21}\sqrt{n_{02}n_{01}} & V_{22}n_{02}
\\\\
V_{11}n_{01} & V_{12}\sqrt{n_{01}n_{02}} &
\frac{q^2}{2m_1}+V_{11}n_{01}+\omega & V_{12}\sqrt{n_{01}n_{02}}
\\\\
V_{21}\sqrt{n_{02}n_{01}} & V_{22}n_{02} &
V_{21}\sqrt{n_{02}n_{01}} & \frac{q^2}{2m_2}+V_{22}n_{02}+\omega
\end{array} \right)
\left( \begin{array}{c} \phi_1 \\\\ \phi_2 \\\\ \phi^*_1 \\\\
\phi^*_2
\end{array} \right)
\end{equation}
\end{widetext}
Solving this gives the excitation spectrum of a two-component CBG
in an external field with arbitrary interactions. It is useful to
make sure the excitation spectrum is the same as that found by
using the BdG equations \cite{alex2comp}. In the presence of no
external fields the determinant is zero and hence gives
\begin{widetext}
\begin{equation}
\omega^2_{1,2}(\bm{k})=\frac{1}{2}\left(
\varepsilon^2_1(\bm{k})+\varepsilon^2_{2}(\bm{k})\pm\sqrt{
[\varepsilon^2_1(\bm{k})-\varepsilon^2_{2}(\bm{k})]^2+
\frac{4k^4}{m_1m_{2}}V^2n_{o1}n_{o2}}\right)
\end{equation}
\end{widetext}
where we assume that $V=V_{11}=V_{22}=V_{12}=V_{21}$.
This result is equivalent to eqn 8. Again this leads to the
acoustic and plasmon modes.

For the purposes of the linear response function, we write the
definitions
\begin{widetext}
\begin{equation}
D_j=-e^*\left( \frac{q^2}{2m_j}\frac{\sqrt{n_{0j}}}{m_j}+
\frac{2V(\bm{q})n_{0j}\sqrt{n_{0j}}}{m_j}  +
\frac{2V(\bm{q})n_{0j'}\sqrt{n_{0j}}}{m_{j'}}
\right)\bm{q}\cdot\bm{a}(\bm{q},\omega)
\end{equation}
\end{widetext}
and
\begin{equation}
B_j=\phi_{j}(\bm{q},\omega)-\phi^*_{j}(-\bm{q},-\omega)
\end{equation}
and solve eqns 17 and 18.
\begin{widetext}
\begin{equation}
\left( \begin{array}{c} D_j \\ D_{j'}
\end{array} \right)=
\left( \begin{array}{cc} a & b
\\
c & d
\end{array} \right)
\left( \begin{array}{c} B_j \\ B_{j'}
\end{array} \right)=\left( \begin{array}{cc}
\omega^2-\frac{q^4}{4m^2_j}-Vn_{0j}\frac{q^2}{m_j} &
-V\sqrt{n_{0j}n_{0j'}}\frac{q^2}{m_{j'}}
\\
-V\sqrt{n_{0j}n_{0j'}}\frac{q^2}{m_{j}} &
\omega^2-\frac{q^4}{4m^2_{j'}}-Vn_{0j'}\frac{q^2}{m_{j'}}
\end{array} \right)
\left( \begin{array}{c} B_j \\ B_{j'}
\end{array} \right)
\end{equation}
\end{widetext}
In the presence of external fields ($\bm{A}\neq 0$) we can solve
for $B_j=\phi_{j}(\bm{q},\omega)-\phi^*_{j}(-\bm{q},-\omega)$:
\begin{eqnarray}
B_j&=&\frac{dD_j-bD_{j'}}{Det\left( \begin{array}{cc} a & b
\\
c & d
\end{array} \right)}
\\
&=&-e^*\frac{(\omega^2-\varepsilon^2_{j'})\frac{\sqrt{n_{0j}}}{m_j}\frac{2m_j}{q^2}
\left( \varepsilon^2_{j}+\frac{Vn_{0j'}q^2}{m_{j'}}
\right)}{(\omega^2-\varepsilon^2_{j})(\omega^2-\varepsilon^2_{j'})-
\frac{V^2n_{0j'}n_{0j}q^4}{m_{j'}m_{j}}}\bm{q}\cdot\bm{a}(\bm{q},\omega)
\nonumber
\\
&-&e^*\frac{\frac{2V\sqrt{n_{0j}}n_{0j'}}{m_{j'}}\left(
\varepsilon^2_{j'}+\frac{Vn_{0j}q^2}{m_{j}}
\right)}{(\omega^2-\varepsilon^2_{j})(\omega^2-\varepsilon^2_{j'})-
\frac{V^2n_{0j'}n_{0j}q^4}{m_{j'}m_{j}}}\bm{q}\cdot\bm{a}(\bm{q},\omega)
\end{eqnarray}
This equation simplifies into the case for a single component
charge Bose gas as found in ref \cite{alexbeere}. This can be done
by taking the second component number density $n_{oj'}\rightarrow
0$:
\begin{eqnarray}
B_j&=&\phi_{j}(\bm{q},\omega)-\phi^*_{j}(-\bm{q},-\omega)
\nonumber
\\
&=&-e^*\frac{\frac{\sqrt{n_{0j}}}{m_j}\frac{2m_j}{q^2}
\varepsilon^2_{j}}
{(\omega^2-\varepsilon^2_{j})}\bm{q}\cdot\bm{a}(\bm{q},\omega)
\end{eqnarray}

Back to the two-component system, the linear-response function
$K^{mn}$ is defined for each component j of the charge Bose gas as
\begin{equation}
J^m_j(\bm{q},\omega)=K^{mn}_j a_n(\bm{q},\omega)
\end{equation}
where $J^m(\bm{q},\omega)$ is the current in the m direction and
$a_n(\bm{q},\omega)$ is the vector potential in the n direction.
The currents of the two components are found to be independent
(from commutation rules of two internal components). Interference
of the two components is taken into account as will be
demonstrated now. For each current we have
\begin{eqnarray}
\bm{J}_j(\bm{r},t)&=&-\frac{ie^*}{2m_j}[\psi^*_{0j}(\bm{r},
t)\bm{\nabla}\psi_{0j}(\bm{r}, t)-(\bm{\nabla}\psi^*_{0j}(\bm{r},
t))\psi_{0j}(\bm{r}, t)] \nonumber
\\
&&-\frac{e^{*2}}{m_j}|\psi_{0j}(\bm{r}, t)|^2\bm{A}(\bm{r},t)
\end{eqnarray}
where as before:
\begin{equation}
\psi_{0j}(\bm{r}, t)=\sqrt{n_{0j}}+\phi_j(\bm{r}, t)
\end{equation}
Keeping terms linear in $\bm{A}$ and fluctuations $\phi_j(\bm{r},
t)$ and performing the Fourier transform we obtain
\begin{eqnarray}
\bm{J}_j(\bm{q},\omega)&=&\frac{e^*\sqrt{n_{0j}}}{2m_j}
[\phi_{j}(\bm{q},\omega)-\phi^*_{j}(-\bm{q},-\omega)]\bm{q}
\nonumber
\\
&&+\frac{e^{*2}n_{0j}}{m_j}\bm{a}(\bm{q},\omega)
\end{eqnarray}
The first term contains $B_j$ and so the current for one component
includes interference effects from the other. Subbing in we obtain
for the current
\begin{widetext}
\begin{equation}
\bm{J}_j(\bm{q},\omega)=\frac{-e^{*2}n_{0j}}{m_j}\left[
\frac{\left[\left\{
\frac{\bm{q}}{q^2}\left(\omega^2-\varepsilon^2_{j'}\right)
\left(\varepsilon^2_{j}+\frac{Vn_{0j'}q^2}{m_{j'}}\right)+
\frac{Vn_{0j'}}{m_{j'}}\bm{q}\left(\varepsilon^2_{j'}+
\frac{Vn_{0j}q^2}{m_{j}}\right)\right\}\bm{q}\cdot\bm{a}
\right]}{(\omega^2-\varepsilon^2_{j})(\omega^2-\varepsilon^2_{j'})-
\frac{V^2n_{0j'}n_{0j}q^4}{m_{j'}m_{j}}}+\bm{a}\right].
\end{equation}
\end{widetext}
We can therefore write the linear-response function as
\begin{widetext}
\begin{equation}
K^{mn}_j = \frac{-e^{*2}n_{0j}}{m_j} \left[\frac{\left\{
\frac{q^mq^n}{q^2}\left(\omega^2-\varepsilon^2_{j'}\right)
\left(\varepsilon^2_{j}+\frac{Vn_{0j'}q^2}{m_{j'}}\right)+
\frac{Vn_{0j'}}{m_{j'}}q^mq^n\left(\varepsilon^2_{j'}+
\frac{Vn_{0j}q^2}{m_{j}}\right)\right\}
}{(\omega^2-\varepsilon^2_{j})(\omega^2-\varepsilon^2_{j'})-
\frac{V^2n_{0j'}n_{0j}q^4}{m_{j'}m_{j}}}+\delta^{mn}\right]
\end{equation}
\end{widetext}
where the linear-response function has been divided into a
longitudinal and transverse part. Since we ultimately want to
examine the dielectric function $\epsilon(\bm{q},\omega)$ we
examine the longitudinal part which can be reduced down to
\begin{equation}
K^{mm}_j = \frac{-e^{*2}n_{0j}}{m_j}
\frac{\omega^2\left[\omega^2-\varepsilon^2_{j'}+\omega^2_{p0j'}\right]}
{(\omega^2-\varepsilon^2_{j})(\omega^2-\varepsilon^2_{j'})-\omega^2_{p0j}\omega^2_{p0j'}}
\end{equation}
where $\omega^2_{p0j}=Vn_{0j}q^2/m_j$.
\begin{figure}[h]
\centering \epsfig{file=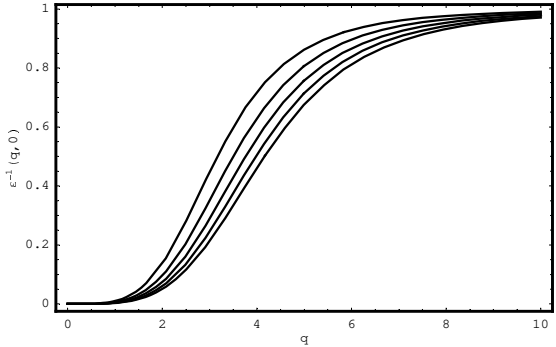, bbllx=99,
bblly=451, bburx=522, bbury=729,width=3in} \caption{The inverse of
the dielectric function for a 2 component charge Bose gas, plotted
with varying density ratio of the 2 components.}
\end{figure}
To obtain the dielectric function we use the continuity equation
\begin{equation}
\bm{\nabla}\cdot\bm{J}=-\frac{d}{dt}\rho_{ind},
\end{equation}
one of the Maxwell equations
\begin{equation}
\bm{\nabla}\cdot\bm{D}=4\pi\rho_{ext},
\end{equation}
the current-external electric field relation
\begin{equation}
\bm{J}(\bm{q},\omega)=\frac{K^{mm}}{i\omega}D(\bm{q},\omega)
\end{equation}
and the usual definition for the dielectric function
\begin{equation}
\frac{1}{\epsilon(\bm{q},\omega)}=\frac{\rho_{tot}}{\rho_{ext}}.
\end{equation}
Here $\rho_{ind}$, $\rho_{ext}$ and $\rho_{tot}$ are induced,
external and total charge density . We find for the inverse
dielectric function
\begin{equation}
\frac{1}{\epsilon(\bm{q},\omega)}=1-\frac{4\pi}{\omega^2}\sum_jK^{mm}_j.
\end{equation}
Doing the summation
\begin{equation}
\frac{1}{\epsilon(\bm{q},\omega)}=1+\frac{\omega^2_{p01}[\omega^2+\omega^2_{p02}-\varepsilon^2_{2}]
+\omega^2_{p02}[\omega^2+\omega^2_{p01}-\varepsilon^2_{1}]}
{(\omega^2-\varepsilon^2_{2})(\omega^2-\varepsilon^2_{2})-\omega^2_{p01}\omega^2_{p02}}
\end{equation}
and rearranging for the static case we have
\begin{equation}
\epsilon(\bm{q},0)=1+\left(\frac{q_s}{q}\right)^4
\end{equation}
where
\begin{equation}
q^4_s=4m^2_1\omega^2_{p01}+4m^2_2\omega^2_{p02}.
\end{equation}
The result is exactly the same as for a single component system
\cite{alexbeere} except here the screening wavenumber is
effectively a sum of number densities. In fig 1 we plot the
inverse static dielectric function. We notice that as the second
components number density is increased the screening becomes
stronger at intermediate q values as expected.

\section{Summary} \label{sect:theseventh}
A two-component CBG with arbitrary interactions has been examined.
The study of its linear response and dielectric function yield
similar results to a single component CBG. The excitation spectrum
in an external field was derived, whilst it was used to re-confirm
previous results which were derived with no external field. This
study of a two-component CBG should be useful for further studies
in bipolaronic systems applied to lattices. Also it will be useful
to work in atomic physics.
\\
\\
This work was funded by EPSRC (grant R46977) and Cambridge
University. We would like to thank A.S. Alexandrov and W.Y. Liang
for their advice and support.


\begin{thebibliography}{10}
\bibitem[1]{alex2comp}  A.S. Alexandrov and V.V. Kabanov, J. Phys.Condens.Matter,
\textbf{14}, L327 (2002).

\bibitem[2]{alexbook}  A.S. Alexandrov, Theory of superconductivity, from
weak to strong coupling, Institute of Physics, Bristol, UK (2003).

\bibitem[3]{alexfree}  A.S. Alexandrov and V.V. Kabanov, Phys. Rev. B, \textbf{59}, 13628 (1999).

\bibitem[4]{hagley}  E.W. Hagley, L. Deng, M. Kozuma,
     J. Wen, K. Helmerson, S.L. Rolston and W.D. Phillips, Science, \textbf{283},
     1706 (1999).

\bibitem[5]{alexbeere}  A.S. Alexandrov and W.H. Beere, Phys. Rev. B,
\textbf{51}, 5887 (1995).

\end{thebibliography}
\end{document}